\documentclass[a4paper]{jpconf}
\usepackage{graphicx}
\usepackage{amssymb}
\usepackage{float}

\begin{document}
\title{Preliminary characterization of a surface electrode Paul trap for frequency metrology}

\author{Josipa Madunic$^{\rm{a}}$, Lucas Groult\footnote{now at Heriot Watt univerity, Edinburgh, Scotland.}, Bachir Achi\footnote{now at Thales AVS/MIS, Vélizy-Villacoublay, France.}, Thomas Lauprêtre\footnote{now at Laboratoire de Physique des Lasers, Université Sorbonne Paris Nord-Institut Galilée, Villetaneuse, France}, Alan Boudrias$^{\rm{a}}$, Pierre Roset$^{\rm{a}}$, Valérie Soumann$^{\rm{a}}$, Yann Kersalé$^{\rm{b}}$, Moustafa Abdel Hafiz$^{\rm{b}}$ and  Clément Lacroûte$^{\rm{a}}$}

\address{$^{\rm{a}}$Université de Franche-Comté, CNRS, institut FEMTO-ST, 26 rue de l’Epitaphe, 25000 Besançon, France\\
$^{\rm{b}}$SUPMICROTECH, CNRS, institut FEMTO-ST, 26 rue de l’Epitaphe, 25000 Besançon, France\\
}

\ead{clement.lacroute@femto-st.fr}

\begin{abstract}

We are developing a single-ion optical clock based on a surface-electrode (SE) trap that we will operate with $^{171}$Yb$^+$ ions on the electric quadrupole transition at 435.5~nm. We present heating rate measurements performed with a prototype SE trap. We also introduce a new, micro-fabricated SE trapping chip using silicon on insulator technology. Electric tests were performed under ultra-high vacuum using a testing chip, including breakdown voltages measurements and flashover detection. We present suitable trapping parameters for this chip, as well as a road-map for improving its design.
\end{abstract}

\section{Introduction}

Atomic clocks are ubiquitous in modern society, and impact various domains such as telecommunications, navigation, energy distribution, etc. Optical atomic clocks are pushing atomic clocks to never-explored territories, reaching systematic fractional uncertainties of $10^{-18}$ and below \cite{huntemann2016, brewer2019, kim2023,srinivas2023}. Such fractional frequency resolution enables a range of new applications, within time and frequency (TF) metrology, fundamental physics \cite{derevianko2016,safronova2018}, geodesy \cite{flury2016, lion2017,mcgrew2018}. Several demonstrations of on-field measurements were performed in the last few years \cite{grotti2018, takamoto2020}, and portable optical clocks are being developed \cite{poli2014,cao2017,koller2017,hannig2019,kong2020, stuhler2021}.

In parallel to the development of optical clocks, trapped-ions based quantum information processing has also made tremendous progress in the past few years, leading to a number of research and industrial achievements \cite{pino2021,auchter2022,zhu2023}. Surface-electrodes (SE) ion traps have been proposed as the base of future quantum computers, with capabilities to control, route and coherently couple hundreds of qubits \cite{chiaverini2005a,seidelin2006}. Technological developments have recently lead to SE ion traps that embedded photonic integrated chips. Such trap use gratings to outcouple light from underneath the trapping electrodes and have been demonstrated with several ionic species  \cite{mehta2016,mehta2020,niffenegger2020}. Such gratings can also be used to collect single-ions fluorescence and are compatible with non-destructive measurements on trapped ions \cite{saha2023}.

We are investigating the use of a SE ion trap in the context of optical frequency metrology \cite{lacroute2016}. We have developed a prototype SE trap, which we characterized experimentally \cite{laupretre2023}. In this article, we summarize the main results of the characterization of this prototype trap and we present an upgraded trapping chip designed and fabricated for optical frequency metrology. We then present initial electric characterization of the chip, including radio frequency (RF) electrodes capacitance and breakdown voltages measurements.

\section{Prototype  surface-electrodes trap}
This section summarizes results obtained with a prototype trap that are presented in more details in \cite{laupretre2023}. A simple, 5-wire geometry SE trap was designed and produced on a printed-circuit board (PCB). It was fabricated externally with commonly used materials to serve as a preliminary test-bed for a future SE trap-based ion clock.

\begin{figure}[h]
\centering
\includegraphics[width=10cm]{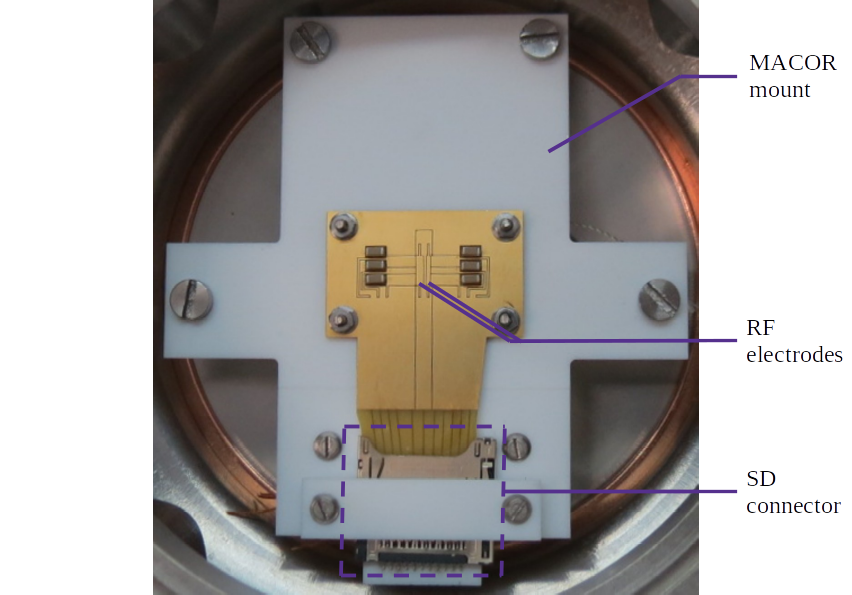}%
\begin{minipage}[b]{14pc}\caption{Photograph of the prototype SE trap used in \cite{laupretre2023}. The trap is held on a MACOR support and electrically connected through a standard SD-connector. It is housed in a vacuum chamber and faces a CF100 window for fluorescence detection.}
\label{fig:mael_trap}
\end{minipage}
\end{figure}

Fig. \ref{fig:mael_trap} shows the prototype trap mounted on its ceramic support,  along with its filtering capacitors and SD-format connections. RF voltages with amplitude $V_{\rm{RF}}$ and frequency $\Omega_{\rm{RF}}$ are applied to the central, rectangular asymmetric electrodes that define the longitudinal axis of the trap. The RF electrodes widths of 600 and 1200~µm induce a so-called pseudo-potential with a trapping distance of about 500~µm. DC voltages are applied to the remaining electrodes in order to close the trap and to cancel DC electric fields at the trap minimum location. The resulting total potential is the combination of the RF pseudo-potential and the DC electrodes potentials, and is close to a harmonic trap with a low longitudinal secular motion frequency and higher transverse secular motion frequencies (see for instance \cite{chiaverini2005a, house2008} for details).

\begin{figure}[h]
\begin{minipage}{14pc}
    \includegraphics[width=8.25cm]{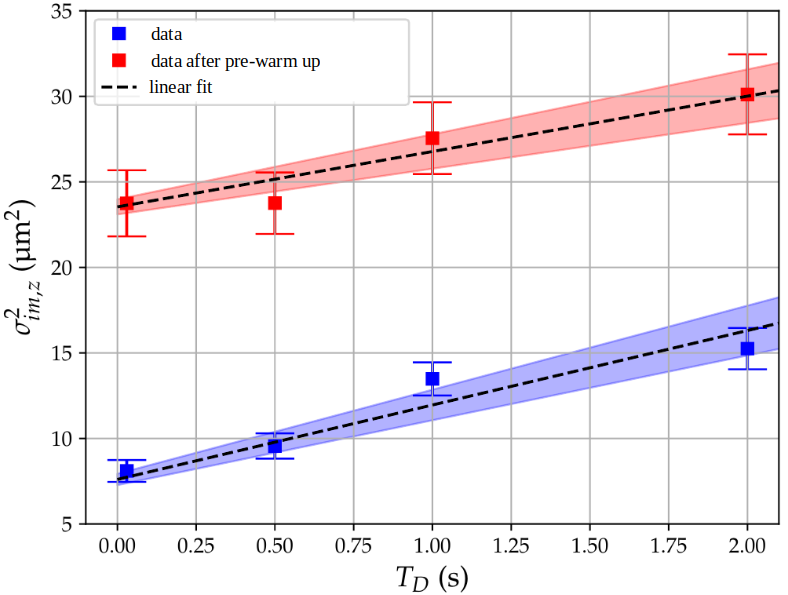}%
\end{minipage}\hspace{5pc}
\begin{minipage}{14pc}
    \includegraphics[width=8.25cm]{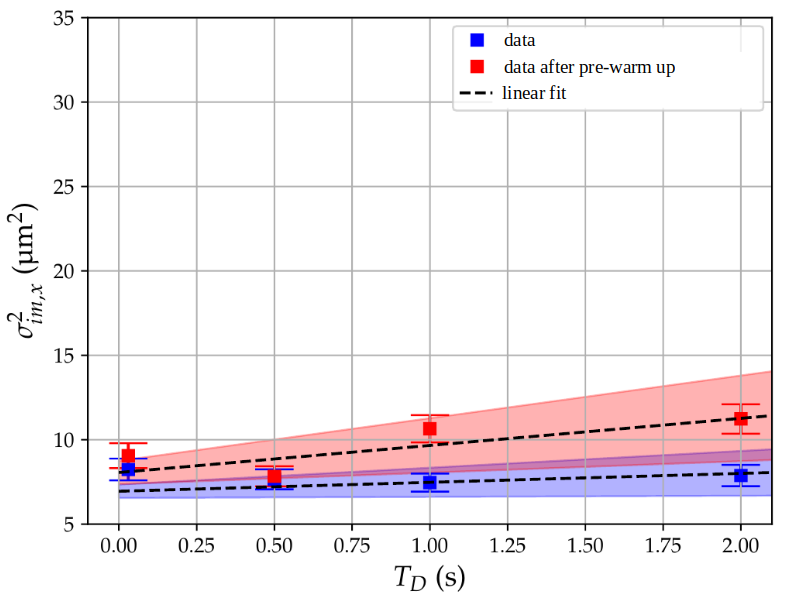}%
\end{minipage}
\caption{Heating rate measurements performed using the prototype chip \cite{laupretre2023}. The ion fluorescence image square width $\sigma^2_{im}$ is plotted versus the dark time $T_D$ along the longitudinal ($x$) and transverse ($z)$ directions. Left: axial heating rate measured with (red) and without (blue) an initial pre-warm up time. Right: transverse heating rate measured with (red) and without (blue) an initial pre-warm up time.}
\label{fig:heating-rate}
\end{figure}

Ions are loaded in the trap from a neutral atom beam generated with a commercial dispenser, using two-photon ionization at 399~nm and 370~nm \cite{laupretre2020}. We obtained an average trap lifetime of 1200~s with cooling lasers on, with trapping frequencies of 85~kHz, 320~kHz and 350~kHz along the longitudinal and transverse directions. The trap heating rate was estimated using a combination of the Doppler-recooling techniques and ion fluorescence imaging. The results obtained by fitting the width of the ion fluorescence spot are shown Fig. \ref{fig:heating-rate}. We observed a heating rate of 8000 phonons/s in all directions, which we think could be related to technical noise on the experiment as well as a non-negligible background pressure at the trap location. Combining the heating rate value with preliminary excess micro-motion measurements, we estimated that a fractional frequency stability of $5{{\times}}10^{-14}\tau^{-1/2}$ and trap-related frequency shifts of order $10^{-16}$ could be obtained with such a trap \cite{laupretre2023}.

This prototype and the related results were the basis of the design of a new trapping chip, with the main objectives being to enhance both the trapping time and heating rate, and to reduce the related fractional-frequency shifts.

\section{New chip design}

The prototype trap used in \cite{laupretre2023} was affected by a number of technical limitations, including:
\begin{itemize}
    \item[--] large inter-electrode spacing (120~µm) that results in exposed dielectrics. These areas can host free charges that distort the trapping potential;
    \item[--] non-ultra-high-vacuum (UHV) dedicated materials that cause a high local pressure in the trap vicinity, reducing the single-ions lifetime via high-energy collisions;
    \item[--] in-plane conductors routes to the connector, which can also distort the trapping potential.
\end{itemize}

We have therefore designed and fabricated a new trapping chip. Silicon is rarely used as a substrate for SE ion traps, because of its high dieletric RF losses. However, doped silicon can be used as an electrode material \cite{britton2009}. Silicon offers many advantages as a material for microfabrication:
\begin{itemize}
    \item[--] it is a streamlined material in any MEMS or CMOS microfabrication facility;
    \item[--] it can easily be etched into high aspect ratios geometries, enabling low inter-electrode spacing and large electrodes height;
    \item[--] it can be used on a number of low-RF losses substrates, including SiO$_2$;
    \item[--] it is inexpensive thanks to its very large use.
\end{itemize}

\subsection{Microfabricated chip geometry and fabrication process}

\begin{figure}[t!]
\includegraphics[width=16cm]{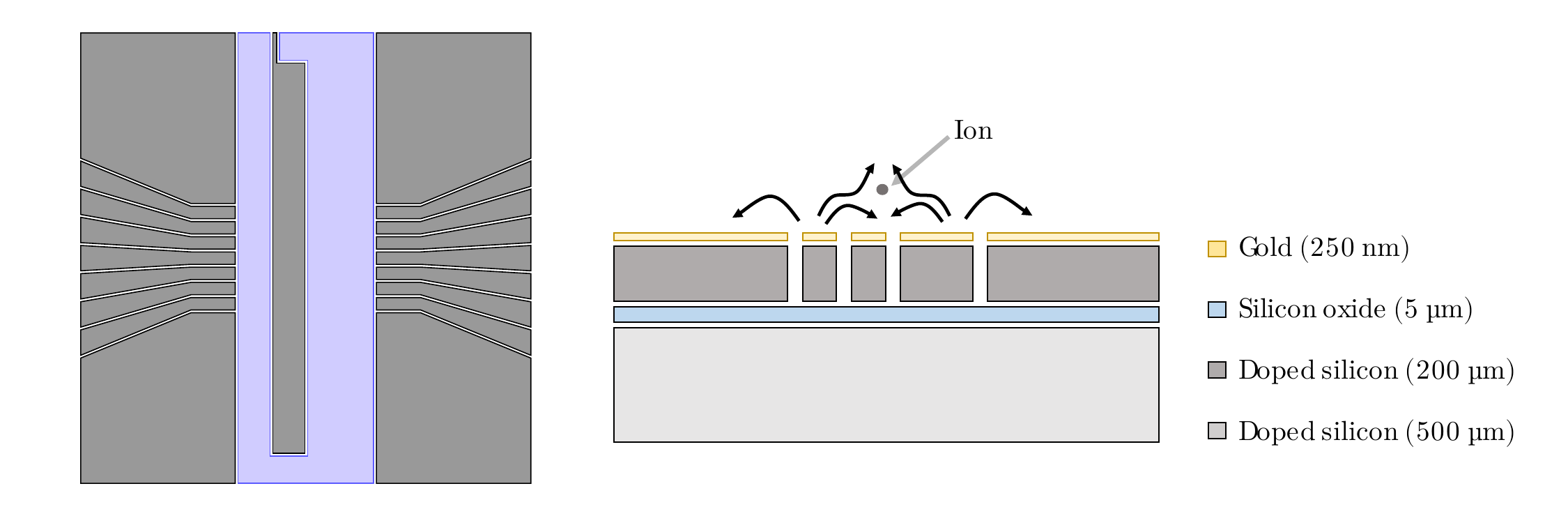}
\caption{Silicon on insulator SE trap design. Left: geometrical scheme of the trapping electrodes. The blue, asymmetric central electrode is the RF electrode. All remaining gray electrodes are DC. Right: transverse cut of the trapping chip. Yellow: gold-coating of the electrodes. Dark gray: doped silicon electrodes. Blue: SiO$_2$ insulation layer. Light gray: doped silicon ground plane. Arrows: electric field lines. An additional 500~µm doped Si layer is not shown, see text for details.}
\label{fig:Lucas_chip}
\end{figure}

We have designed a trapping geometry based on a 5-wire geometry similar to the prototype trap, shown Fig. \ref{fig:Lucas_chip}. With 600~µm and 1200~µm wide RF electrodes, the trapping distance of 500~µm is unchanged. We added control electrodes, which enable finer displacement of the trap location in the longitudinal direction, as well as the generation of multiple trap locations. This opens the possibility of multi-ions experiments.

The chip has been fabricated using deeply reactive ion etching (DRIE) on a silicon on insulator (SOI) wafer. The initial wafer consists of a 200~µm-thick doped-silicon device layer, a 5~µm-thick buried oxyde (box) SiO$_2$ layer and a 500~µm-thick handle layer (see Fig. \ref{fig:Lucas_chip}). The handle is doped in order to serve as an electrical ground plane.

In order to increase the chip overall height so that its surface rises above the chip carrier, the wafer handle side is bonded to a 500~µm thick doped silicon wafer. This results in a total 1.2~mm chip height, which raises the electrodes surfaces about 100~µm above the chip carrier surface (commercial ceramic pin grid array).

The electrodes are gold coated, and etched through the device layer using DRIE, resulting in 200~µm high electrodes with 20~µm inter-electrodes distance. The resulting 1/10 aspect ratio ensures that the ion is shielded from the box by the electrodes.

\begin{figure}[h!]
\centering
\includegraphics[width=12pc]{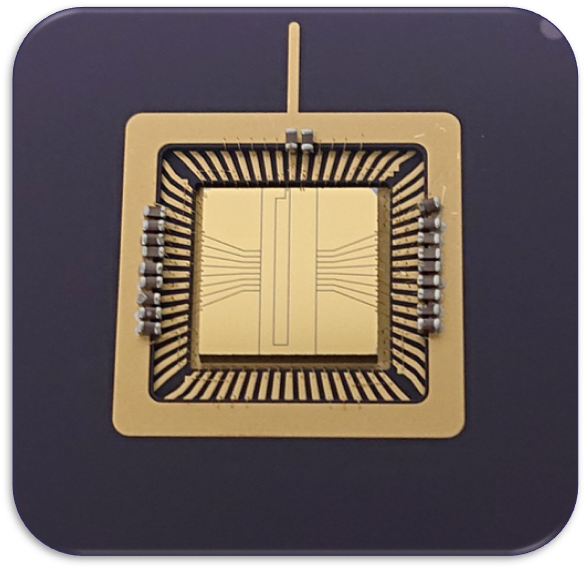}\hspace{2pc}%
\begin{minipage}[b]{14pc}\caption{5-wire surface-electrode trap microfabricated at FEMTO-ST. The trap is embedded in a commercial, UHV compatible ceramic chip carrier. Filter capacitors prevent RF leakage to the DC electrodes.}
\label{fig:chip_photo}
\end{minipage}
\end{figure}

Fig. \ref{fig:chip_photo} shows a picture of the finalized microfabricated chip. It is glued inside a ceramic pin grid array carrier (CPGA) with conductive epoxy. Electrodes are wire-bonded to the CPGA pads. 220 nF capacitors are connecting all DC electrodes to ground in order to prevent RF leakage from the RF to the DC electrodes. The chip is a 8-mm long square, while the CPGA length is one inch. The overall size of the setup will be determined by the vacuum chamber volume as well as the optical setup and control electronics. Using traditional components, an overall size of a few hundred liters can be achieved, see for instance \cite{stuhler2021}. Using dedicated photonic integrated chips, digital electronics \cite{pruttivarasin2015} and a compact vacuum cell \cite{wilpers2013}, this could be reduced to the liter scale.%

\subsection{Trapping parameters}
Trapping parameters were derived to operate the clock in a Doppler-free regime. This can be ensured by operating the clock laser and ion trap in the resolved-sideband, recoil-free regime \cite{ludlow2015}. The resolved-sideband condition is $\Gamma\ll\omega$, where $\Gamma$ is the clock transition width and $\omega$ is the secular motional frequency along the laser direction. Recoil-free clock laser interrogation is ensured when $E_{\rm{recoil}}\ll\hbar\omega$ where $E_{\rm{recoil}}=\hbar\omega_{\rm{recoil}}$ is the recoil energy associated with the clock laser photons. For the quadrupole transition of Yb$^+$, $\Gamma\approx2\pi{\times}3$~Hz and $\omega_{\rm{recoil}}\approx2\pi{\times}6$~kHz; for typical radial trapping frequencies $\omega\gtrsim2\pi{\times}100$~kHz, both the resolved-sideband regime and recoil-free interrogation are ensured. 


We devised two sets of trapping parameters, to be able to reach either $\omega>2\pi{\times}100$~kHz or $\omega\geq2\pi{{\times}}500$~kHz. These parameters are given in table \ref{table:trap}. Both would enable working in the resolved-sideband regime. However, higher secular frequencies are preferred in order to reduce the decoherence caused by heating during interrogation \cite{laupretre2023}. Set (1) would therefore be more favorable in that regard, but as shown in the next section, the required RF voltage might be hard to reach. Set (2) represents a compromise to reach the Doppler-free regime with more realistic voltages. 

\begin{table}[h]
\caption{\label{table:trap}Proposed trapping parameters. $\Omega_{\rm{RF}}$ is the trap RF drive frequency, $V_{\rm{RF}}$ is the RF voltage amplitude, $V_{\rm{DC}}^{\rm{max}}$ max indicates the maximum required DC voltage, $q$ is the transverse RF stability parameter.}
\begin{center}
\begin{tabular}{llllllll}
\br
Set & $\Omega_{\rm{RF}}$ & $V_{\rm{RF}}$ & $V_{\rm{DC}}^{\rm{max}}$  & $\omega$ & Trap depth & $q$ & $\eta$ \\
\mr
1 & 4.9 MHz & 300 V & 7 V & 500 kHz & 325 meV  & 0.28 & 0.11 \\ 
2 & 3 MHz   & 100 V & 7 V & 310 kHz & 95 meV & 0.26 & 0.14 \\ 
\br
\end{tabular}
\end{center}
\end{table}

\section{New chip preliminary tests results}
In order to trap single ions, we apply high radiofrequency voltages to the RF electrode using a simple circuit. This circuit forms an RLC resonator with the chip, which is decoupled from the source by a transformer. The trap voltage is read on a capacitor bridge with a division ratio of 100. The loaded quality factor is measured as the ratio $f_0/\Delta f$, where $f_0$ is the resonant frequency and $\Delta f$ is the frequency interval where the voltage transmission is greater than the maximum over $\sqrt{2}$. Values in the range of 60-70 are usually measured measured.

\subsection{Electrical characterization}

Determining the capacitive and resistive loads formed by the trapping electrodes within an ion trap system, represented as an RC circuit, is crucial for constructing a highly accurate experiment. Determining the capacitance of the chip enables us to set the value of the resonator inductance for the desired RF frequency $\Omega_{\rm{RF}}$. It also gives information on the amount of power dissipated by the resonator and the heat generated in the process. The value of resistance sets the quality factor and RF voltage gain across the resonator, and also contributes to the overall power dissipation.\\ 

The presented trapping chip is modeled as a capacitor and resistor in series, with typical measured values of 80 to 100 pF for capacitance and around 3 $\Omega$ for resistance. These values were determined through a series of measurements using a vector network analyzer (VNA) on 3 different chips. A calibration of the VNA is carried out for each measurement. In this setup, the VNA measures the S$_{11}$ parameters. These parameters were converted to express the capacitance and resistance of the chip between the RF electrode and ground plane. 
The dissipated power in the resonator is given by: $P=(C\cdot\Omega_{\rm{RF}}\cdot V_{RMS})^2\cdot R$ \cite{marcus_d_hughes_microfabricated_2011}. Assuming a total resistance of 3 $\Omega$, a trap capacitance of 100 pF, an RF trap frequency of 4 MHz and a voltage of 200 V amplitude, the dissipated RF power is around 0.4 W, which is significant as compared to other SE trap-based resonators in the literature and could lead to heating of the chip. Indeed, when the trap is operated with a voltage amplitude of 200 V, an increase in pressure at the level of $10^{-12}$  mbar is observed. \\ 
Table \ref{table:contributions} summarizes the contributions to the overall capacitance and resistance of various components, including the chip, wire bonding and inductor in the setup, for one of the tested devices, with resonant frequency of 3.9~MHz. All values were measured using the VNA except for the total quality factor which was directly measured as $\frac{f_0}{\Delta f}$. The quality factor of the chip and inductor have been calculated from the measured values of resistance, capacitance and inductance. 

\begin{table}[h!]
\caption{\label{table:contributions}Overview of different contributions to overall capacitance and resistance. Capacitances, resistances and quality factor of the resonance are given in the table. The capacitance of the wires and inductors are considered to be negligibly small compared to the capacitance of the chip. }
\begin{center}
\begin{tabular}{lccccc}
\br
 & Chip & Wire bonding and carrier & Inductor & RF source & Total\\
\mr
Capacitance (pF) & 107 & - & - & - & 107   \\ 
Resistance ($\Omega$) & 3   & 0.6 & 0.5 & 2.2 & 6.3   \\ 
Q & $127$   & na & 312 & na & 60   \\ 
\br
\end{tabular}
\end{center}
\end{table}



The directly measured $Q$ value almost perfectly matches the value calculated from the losses measured by the VNA $Q=1/(2\pi f_0CR)$. The main contribution to the losses comes from the trapping chip itself. This is mostly explained by the rather low conductivity of doped silicon compared to metals. We are currently investigating ways to enhance this design and reduce both ohmic and capacitive losses, which can be performed with two main changes:
\begin{itemize}
    \item[--] increasing the insulator layer thickness would decrease the chip capacitance, which would result in a reduction in power dissipation and an increase in RF voltage breakdown value.
    \item[--] reducing the total resistance of the doped silicon elements, using wafers with the highest possible conductivity as well as by optimizing the electrodes geometry. This will necessitate finite-element modeling of the chip in order to account for RF ohmic losses in the trap.
\end{itemize}

\subsection{Breakdown voltage tests}
SE traps require high RF voltages, mainly dictated by the mass of the ion, electrodes configuration and the desired trapping frequencies \cite{mcloughlin2011,hong_guidelines_2016}.
However excessive voltages applied to the electrodes can lead to RF breakdown \cite{bautista-salvador_multilayer_2019,sterling_increased_2013}. The impact of the breakdown damage to the surface of the electrode depends on the location, scale, and physical structure of the damage.\\
Concerns arise in the design of micrometer-scale ion traps due to the relatively lower breakdown threshold at RF frequency \cite{stick_ion_2006}. This issue is particularly significant for surface-electrode ion traps, which exhibit a shallower trap depth compared to conventional linear Paul traps. The need for a deep well depth and high trapping efficiency is critical. Hence, the design of microtrap structures that allow for a high applied voltages is a favorable condition. This requires a high inter-electrode separation.\\
Breakdown can also occur between the RF electrodes and the ground plane. In surface ion traps, where high RF voltages between the electrodes and the ground plane are desirable, the dielectric layer separating the two metal layers should be thick to maximize the breakdown voltage. In our case, the thickness of the dielectric layer (SiO$_2$) is 5 µm, limited by the commercially available SOI wafer, which adds constraints to the voltages that can be applied to the electrodes. This limitation makes achieving ion trapping with the desired stability parameters challenging.\\

\begin{figure}[h]
\centering
\includegraphics[width=40pc]{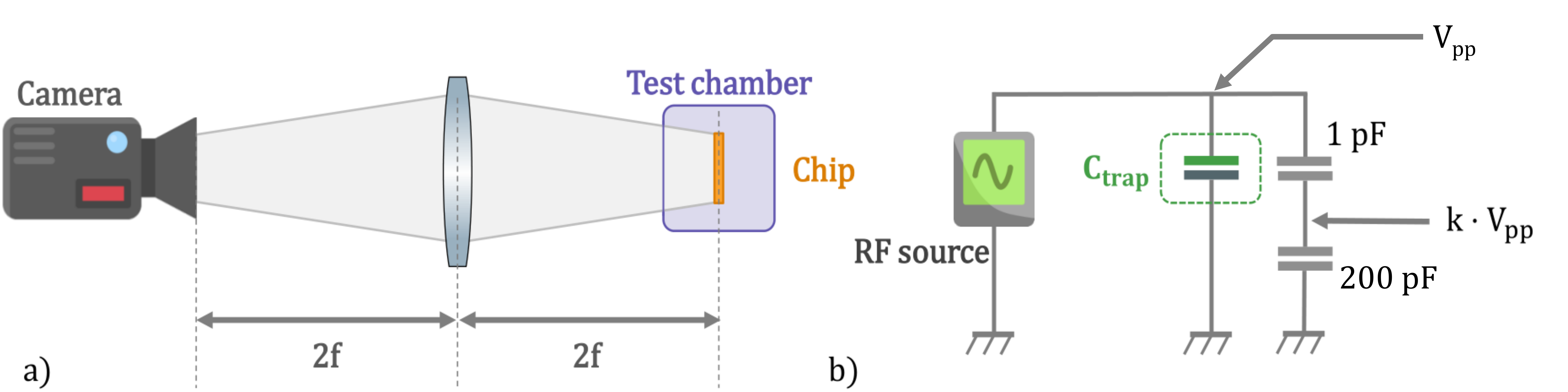}
\caption{\label{label} a) The chip is placed in the test chamber at a pressure of $6 \times 10^{-11}$ mbar. Throughout the test, a voltage ramp is applied, and data acquisition is carried out using a CCD camera. b) Experimental setup schematic for supplying and reading the trap voltage with  the chip in a vacuum chamber. The voltage is supplied using an RLC circuit resonator and monitored through a capacitor bridge with a ratio of k=1/200 with respect to the trap voltage V$_ {pp}$.}
\label{fig:setup_flash}
\end{figure}

In order to measure the typical voltage the RF electrodes can handle without breakdown, several flashover tests were conducted in a test vacuum chamber with pressures lower than $6\times 10^{-11}$ mbar. The experimental setup is shown in Fig. \ref{fig:setup_flash}.  A 1:1 imaging system, with a CCD camera, was used to observe the surface of the chips. The camera, cooled to -80°C, captures images with a 3-second acquisition time every 4.07 seconds.\\

\begin{figure}[h!]
\begin{minipage}{11pc}
    \includegraphics[width=5cm]{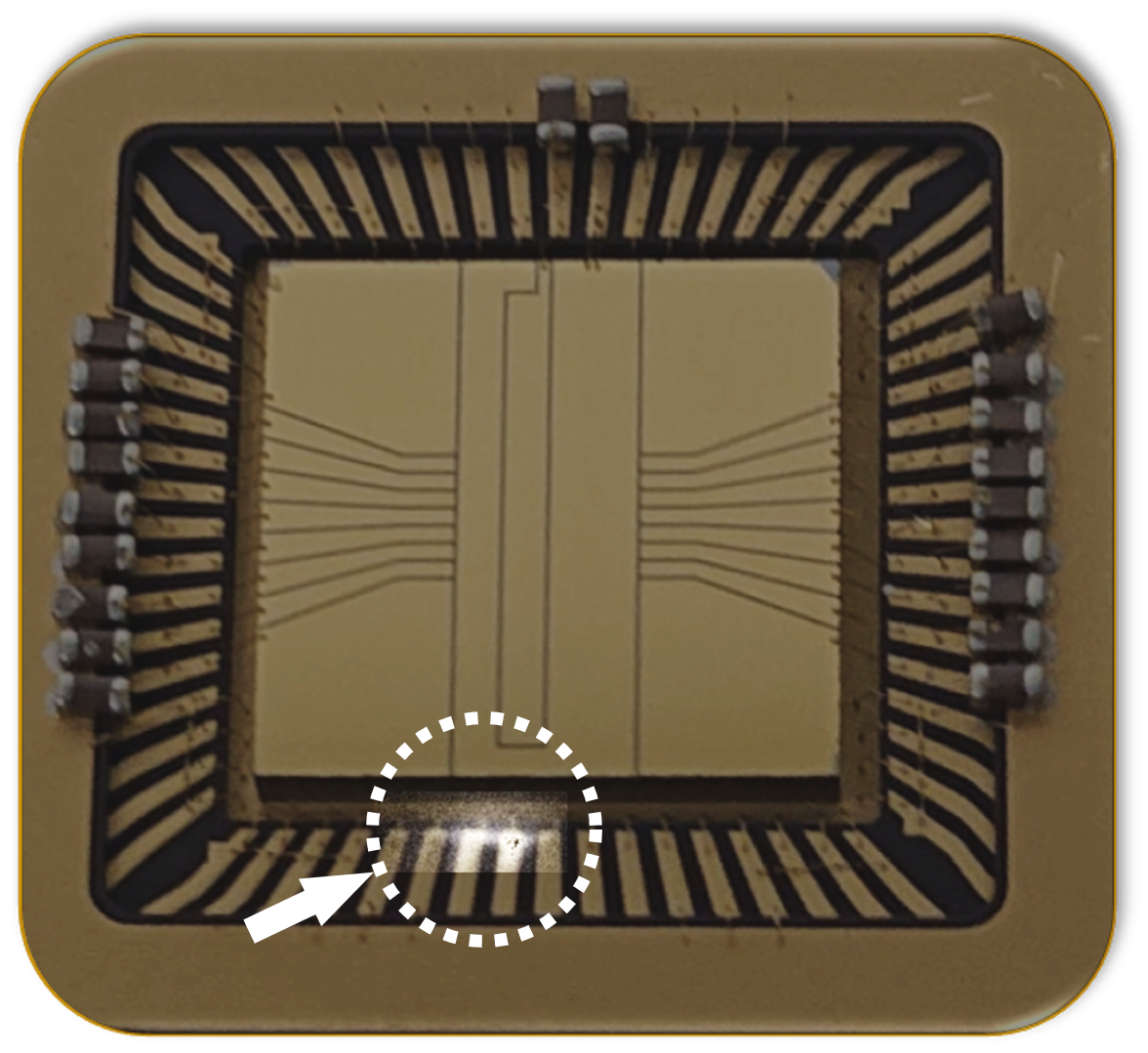}%
\end{minipage}\hspace{1pc}
\begin{minipage}{14pc}
    \includegraphics[width=11cm]{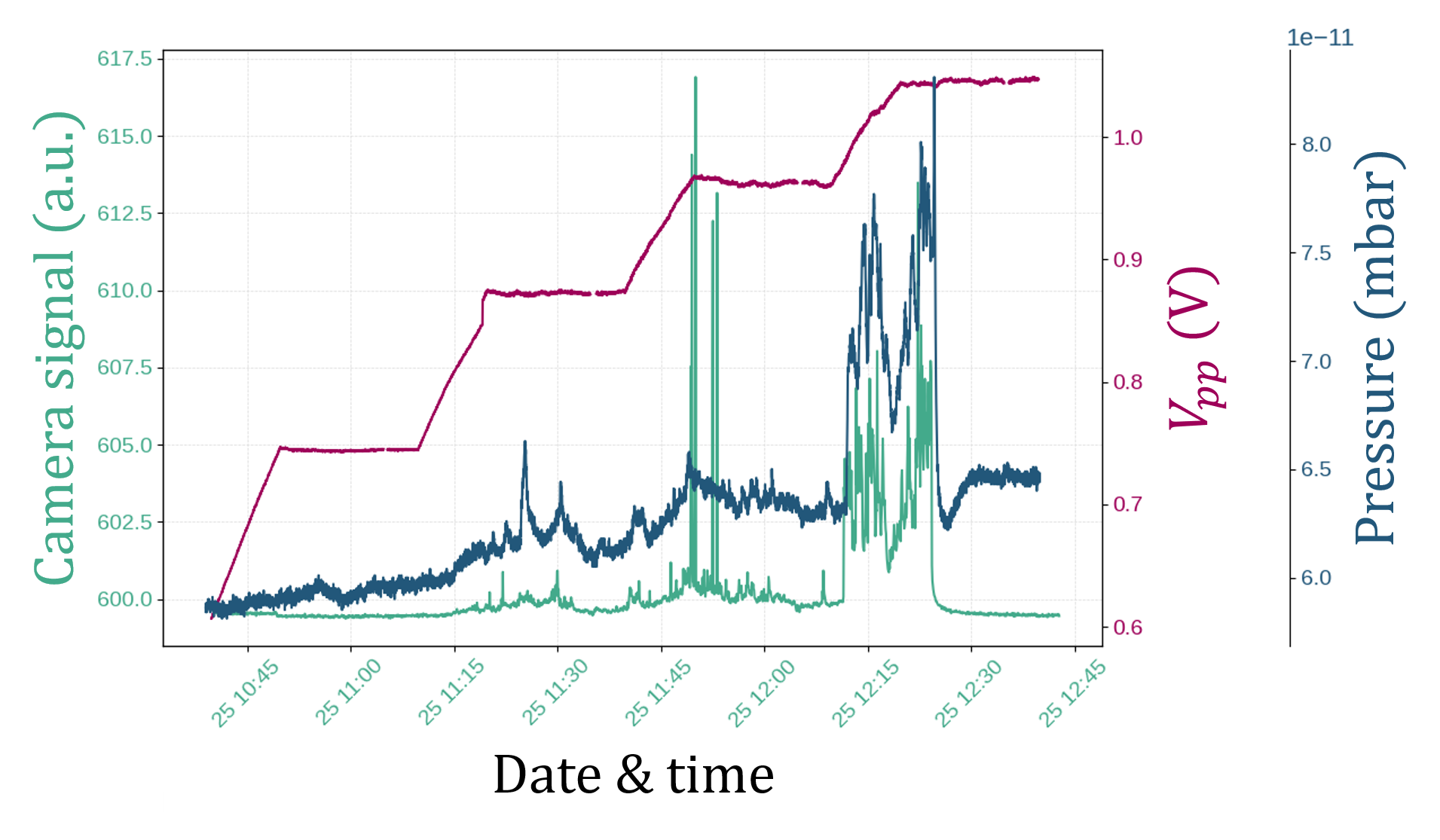}\hspace{1pc}%
\end{minipage}
\caption{Left: Depiction of the chip surface with the reflection from the flashover highlighted on the chip carrier pads within a white circle. Right: Corresponding flashover data. The camera signal integrated over a region of interest centered on the flashing region is represented in green, pressure data in blue, and the voltage read on the monitor in pink.}
\label{fig:Breakdown_results}
\end{figure}

The measurement was conducted for a duration of two hours. 10 minutes voltage ramps followed by a 20-minute wait at the resonant frequency and a quality factor measurement are applied to the RF electrode. From 150 V peak-to-peak on the RF electrode, reflections of flashes occurring at the edge of the chip started to be observed (see Fig. \ref{fig:Breakdown_results}, left). However, voltages of several hundred volts peak-to-peak were reached without clear breakdown observed on the voltage monitor, Fig. \ref{fig:Breakdown_results}, right. Flashing events observed on the camera signal were accompanied by a pressure increase of around $10^{-11} $~mbar, as shown in Fig. \ref{fig:Breakdown_results}, right. After this voltage was reached, flashes are visible in nearly all acquired images, and their intensity increases with higher voltage. They stopped spontaneously before the end of the experiment, at 200~V peak-to-peak (Fig. \ref{fig:Breakdown_results}, right).\\

Following the conclusion of these experiments, the trapping parameters specified in Table \ref{table:trap} were formulated. The first configuration of parameters is not achievable with the existing design; however, the second configuration might be possible.\\
After the measurements, the chip was inspected using the SEM microscope, and no clearly visible damage was observed on the electrodes at the location of the flashing events. The results of these tests have shown that flashover is most likely to occur at voltages of 150 V peak-to-peak. The obtained results will guide the design of an upgraded version of the chip, with a thicker layer of insulator between the electrodes and the ground plane.

\section{Conclusion}
We are investigating the use of silicon-based SE traps for optical frequency metrology. Using a SE trap combines the advantages of offering a very reduced footprint and the possibility to integrate optical as well as electrical functions. Silicon is the material of choice for micro-fabrication, and the realization of silicon-based, high-performance SE traps for optical atomic clocks would be a key technological advance towards industrial, transportable optical atomic clocks.

In this article, we have summarized experiments performed with a prototype SE trap, and presented a new micro-fabricated trapping chip that underwent initial electric characterization. Further improvements in the design are already identified and planned for reaching higher values of $V_{\rm{RF}}$ and $\omega$. This new chip is compatible with UHV pressures and trapping configurations in the resolved-sidebands regime. Future experiments will include heating rate measurements as well as coherence time measurements on the optical clock transition at 435.5~nm, for integration in a transportable ion clock.

\section*{Acknowledgments}
The authors would like to thank Philippe Abbé and Yannick Gruson from the TF department at FEMTO-ST for their technical help with chip connections and support and RF management. We would also like to thank Nicolas Passilly from the MN2S department and William Daniau from the TF department at FEMTO-ST for their advice on the chip microfabrication process.

This work has been supported by the EIPHI Graduate School (contract “ANR-17-EURE-0002”), by the ANR-10-LABX-48-01 First-TF and ANR-11-EQPX-0033 Oscillator-IMP, by the Région Bourgogne Franche-Comté and by the French RENATECH network and its FEMTO-ST technological facility.



\section*{References}
\bibliographystyle{iopart-num}
\bibliography{FSM2023}
\end{document}